# Effective Hamiltonian for piecewise flat potentials and masses

Liès Dekar
Institut des Sciences et de la Technologie
Centre Universitaire de Médéa, DZ 26000 Médéa, Algeria

We consider a class of Hermitian Hamiltonians with position-dependent mass $H = (m^\alpha\, p\, m^\beta\, p\, m^\alpha)/2 + V$ with $2\alpha + \beta = -1$. We apply these Hamiltonians to different piecewise flat potentials and masses (step, barrier, well and multibarrier). To raise the ordering ambiguity we impose that the transmission coefficient tends to the unity as the energy increases indefinitely. We arrive at the conclusion that the form $H_{flat} = (m^{-1/4}\, p\, m^{-1/2}\, p\, m^{-1/4})/2 + V$ of the effective-mass Hamiltonian is the most adequate to describe such flat potentials and masses systems.

Email address: ldekar@yahoo.fr

## I. INTRODUCTION

The recent crystallographic growth techniques have become so fine that they allow the production of non uniform semiconductors with practically abrupt heterojunctions. The understanding of transport properties through these semiconducting heterostructures is indispensable for the prediction of the performances of these new samples. In these mesoscopic materials, the effective mass of the charge carriers, which depends on the type of crossed material, are position dependent [1]. The problem of the choice of an effective Hamiltonian, suitable to govern this kind of systems with position dependent mass, is then posed. Indeed, since the mass and momentum operators no longer commute, there are, a priori, several manners to order these two factors in the kinetic energy operator *T*. Different forms of *T* are proposed in the literature. The majority results from the Hermitian form established by von Roos [2]

$$T(\alpha, \beta, \gamma) = (m^\alpha\, p\, m^\beta\, p\, m^\gamma + m^\gamma\, p\, m^\beta\, p\, m^\alpha)/4 \qquad (1)$$

where $\alpha + \beta + \gamma = -1$.

For instance, Gora and Williams [3] have used $T(-1,0,0) = (m^{-1}\, p^2 + p^2\, m^{-1})/4$, Zhu and Kroemer [4] have opted for $T(-1/2, 0, -1/2) = (m^{-1/2}\, p^2\, m^{-1/2})/2$, Ben Daniel and Duke [5] and more recently Lévy-Leblond [6] and Dekar [7] have proposed

$$T(0,-1,0) = (p\, m^{-1} p)/2, \tag{2}$$

Li and Kuhn [8] have formulated the so-called "redistributed model" $T(0,-1/2,-1/2) = (p\, m^{-1/2} p\, m^{-1/2} + m^{-1/2} p\, m^{-1/2} p)/4$.

The majority of applications, considered by these authors, have been made for piecewise constant potentials and masses. It is also interesting to study systems with continuously variable potential and mass; this was dealt within a previous work [9], where, using the kinetic energy operator (2), we proposed a solution of the Schrödinger equation for a smooth potential and mass step.

Breaking the ambiguity upon the choice of the α and β parameters is not obvious. In a previous paper [7], we have opted for the following strategy: using the generalized kinetic operator (1), and after computing the transmission coefficient for a system with smooth potential and mass step, we proceed to the passage to the limiting case of an abrupt potential and mass step. The conclusion of this work was that the parameters α, β, and γ must take the values α = γ = 0, and β = -1. This is in agreement with the Lévy-Leblond result [6], where the notion of the Galilean instantaneous invariance has been used.

In the present work, we propose to specify the α and β values for one-dimensional systems with piecewise constant potentials and masses. This kind of flat potentials and masses are often used, in first approximation, to describe transport properties in semiconducting heterostructures. Indeed, as a first approximation, we can consider that the carrier propagates through constant potentials - determined by the band structure of each material - provided the mass of the carrier is substituted by the effective mass, which, of course, depends on the crossed material.

We have, in a precedent work [11], already used this kind of flat potential and mass, more precisely the cases of an abrupt step, an abrupt rectangular barrier and a multibarrier, to compute, via path integral formalism, their Green's functions resulting from the kinetic energy operator (2). The transmission coefficients induced by these Green's functions are exactly equal to those found by Lévy-Leblond [12], who has solved the Schrödinger equation with the same kinetic operator (2). Comparatively to the standard case with constant mass, the curves of the transmission coefficient, versus the energy, reveal new behaviors. In particular, for the case of an abrupt potential and mass step, the transmission coefficient reaches the unity for a certain value of the energy, and, as the energy tends to the infinity, the transmission coefficient tends to a limit which is lower than the unity. For a rectangular potential and mass barrier, the transmission coefficient reaches the unity once more than the usual "transparencies" of Ramsauer-Towsend. Moreover, as the energy tends to the infinity, the transmission coefficient does not tend to the unity, but continues to oscillate between the unity and a minimal asymptotic value [11, 12]. For the potential and mass multibarrier, as the energy increases, the width of the forbidden energy bands shrinks more slowly than for the case of a constant mass [12].

For an abrupt step and barrier, the fact that the transmission coefficient does not tend to the unity when the energy of the carriers increases, is, at least, surprising. Sassoli de Bianchi and Di Ventra [13] have showed, for the case of rectangular potential and mass barrier, and using the kinetic operator (2), that the transmission coefficient tends to the unity as the energy increases indefinitely, provided that the mass is a continuous function of position.

We are in a position to ask the following question: is the choice of the kinematic operator and/or the discontinuity of the effective mass that induce this strange asymptotic behavior of the transmission coefficient? To answer this question, we have, in this present work, adopted the following procedure: firstly, we have applied the connection rules on the wave function and its derivative, established by Morrow and Browstein [14], namely, the continuity across interfaces of:

$$\begin{cases} m^\alpha \Psi \\ m^{\alpha+\beta}\, d\Psi/dx \end{cases}. \qquad (3)$$

These authors have showed that in (1) we must have $\alpha = \gamma$, otherwise the heterojonction behaves like an impenetrable barrier which is, evidently, a nonphysical result. The effective Hamiltonian resulting from (1) with the restriction $\alpha = \gamma$:

$$H = \left(m^\alpha\, p\, m^\beta\, p\, m^\alpha\right)/2 + V, \qquad (4)$$

where $2\alpha + \beta = -1$, implies the connection rules (3). Secondly, we impose, in an ad hoc way, the condition that the transmission coefficient tends to the unity as the energy tends to the infinity, which seems to be thoroughly natural condition. Then, we investigate the involvements on the parameter β of the application of these conditions, in the cases of an abrupt potential and mass step in Sec. II, an abrupt rectangular potential and mass barrier (resp. well) in Sec. III, and a potential and mass multibarrier in Sec. IV. In the conclusion, given in Sec. V, we will see that the $\beta = -1/2$ choice, i.e., a kinetic energy operator equal to:

$$T(-1/4,-1/2,-1/4) = \left(m^{-1/4}\, p\, m^{-1/2}\, p\, m^{-1/4}\right)/2, \qquad (5)$$

secures that, as the energy increases, the transmission coefficient, for piecewise potentials and masses, tends to the unity. Furthermore, this choice, de facto, eliminates all the conflicting behaviors with regards to the standard case of a constant mass.

## II. ABRUPT POTENTIAL AND MASS STEP

We assume that the potential and the mass present a discontinuity of the first order at $x = 0$ position. They are given by:

$$V(x) = V_0\, \Theta(x), \qquad (6)$$

$$m(x) = m_1 + (m_2 - m_1)\, \Theta(x), \qquad (7)$$

where $\Theta(x)$ is the Heavyside step function. Using the kinetic energy operator (4), the resulting Schrödinger equation is:

$$-\frac{\hbar^2}{2} m^\alpha \frac{d}{dx}\left[m^\beta \frac{d}{dx} m^\alpha \Psi(x)\right] + [V(x) - E]\Psi(x) = 0. \tag{8}$$

Putting $k_1^2 = 2m_1 E/\hbar^2$ and $k_2^2 = 2m_2(E - V_0)/\hbar^2$, and assuming that the particle is coming from the left, the time independent wave function takes the form:

$$\Psi(x) = \begin{cases} A\exp(ik_1 x) + B\exp(-ik_1 x) & (x<0) \\ C\exp(ik_2 x) & (x>0) \end{cases}. \tag{9}$$

The connection rules (3) at $x = 0$ give:

$$\begin{cases} m_1^\alpha (A+B) = m_2^\alpha C \\ k_1 m_1^{\alpha+\beta}(A-B) = k_2 m_2^{\alpha+\beta} C \end{cases} \tag{10}$$

The transmission coefficient is then given by:

$$\mathfrak{I}_{A.S} = (k_1/m_1)(m_2/k_2)|C/A|^2 \tag{11}$$

Solving the system (10), and putting $\sigma = (m_1/m_2)^{\beta+1/2}$ we find for $\mathfrak{I}_{A.S}$:

$$\mathfrak{I}_{A.S} = \frac{4\sigma\left|\sqrt{E(E-V_0)}\right|}{\left|\sigma\sqrt{E} + \sqrt{E-V_0}\right|^2} \tag{12}$$

This corresponds to a reflection coefficient equal to:

$$\mathfrak{R}_{A.S} = \left|\frac{\sigma\sqrt{E} - \sqrt{E-V_0}}{\sigma\sqrt{E} + \sqrt{E-V_0}}\right|^2. \tag{13}$$

This last expression shows that if $E \langle V_0$ we have $\mathfrak{R}_{A.S} = 1$, i.e., there is total reflection. At this stage, as the energy tends to the infinity, we impose that the limit of (12) must be equal to the unity:

$$\lim_{E\to\infty} \mathfrak{I}_{A.S} = 4\sigma/(\sigma+1)^2 = 1. \tag{14}$$

This implies $\sigma = 1$, i.e., the $\beta$ parameter must necessarily be equal to $-1/2$.

The finite value $E_t(\beta)$ of $E$ for which the potential and mass step becomes transparent, i.e. $\Im_{A.S} = 1$, is :

$$E_t(\beta) = V_0 / (1 - \sigma^2) . \quad (15)$$

Since $E_t(\beta)$ must be greater than $V_0$, we must have $\sigma^2 < 1$. We have three possible cases: (i) $m_1/m_2 > 1$, which implies $\beta < -1/2$, (ii) $m_1/m_2 < 1$, gives $\beta > -1/2$, and (iii) $\beta = -1/2$, implies that $E = E - V_0$, which is impossible. So, if we choose $\beta = -1/2$, there is no value of the energy which cancels the reflection coefficient. We can imagine an experiment where we detect this value $E_t(\beta)$ and, knowing the $m_1/m_2$ ratio, an indication about $\beta$ can be provided. In return, if $E_t(\beta)$ does not exist, this will prove that $\beta = -1/2$ is the right choice.

## III. ABRUPT RECTANGULAR POTENTIAL AND MASS BARRIER (WELL)

Assuming that the mass and the potential have discontinuities at the same positions, we have:

$$V(x) = \begin{cases} 0 & (x < 0, x > a) \\ V_0 & (0 < x < a) \end{cases} \quad (16)$$

$$m(x) = \begin{cases} m_1 & (x < 0, x > a) \\ m_2 & (0 < x < a) \end{cases} \quad (17)$$

The Schrödinger equation (8) gives for the wave function of particle incoming from the left:

$$\Psi(x) = \begin{cases} A \exp(i k_1 x) + B \exp(-i k_1 x) & (x < 0) \\ C \exp(i k_2 x) + D \exp(-i k_2 x) & (0 < x < a) \\ F \exp(i k_1 x) & (x > a) \end{cases} \quad (18)$$

Applying the connection rules (3) at $x = 0$ and $x = a$, we obtain for the transmission coefficient:

$$\Im_{A.B} = \left(1 + g(E,\beta) \sin^2 k_2 a\right)^{-1}, \quad (19)$$

with:

$$g(E,\beta) = \frac{[(\sigma^2 - 1)E + V_0]^2}{4\sigma^2 E(E - V_0)} . \quad (20)$$

The limit of $g(E,\beta)$, as $E$ tends to the infinity, is :

$$\lim_{E \to \infty} g(E,\beta) = (\sigma^2 - 1)/4\sigma^2 .\tag{21}$$

If we impose that $\lim_{E \to \infty} \Im_{A.B} = 1$, i.e. $\lim_{E \to \infty} g(E,\beta) = 0$, then we must have $\beta = -1/2$. In this case, the transmission coefficient is identical to that of a constant mass equal to $m_2$.

It is remarkable that the transmission coefficient reaches unity, except for the usual "transparencies" of the Ramsauer-Towsend type, for the same value $E_t(\beta)$ (18) of the energy obtained in the case of abrupt step. We can see, here again, that if $\beta = -1/2$, $E_t(\beta)$ does not exists. Concurrently to the abrupt step case, we can imagine an experiment where we can detect this value $E_t(\beta)$, and then draw the same conclusions.

In the $0 \langle E \langle V_0$ case, and if we consider the thick barrier approximation, i.e., if $\kappa_2\, a \langle\langle 1$, where $\kappa_2^2 = 2\,m_2\,(E - V_0)/\hbar^2$, the transmission coefficient is:

$$\Im_{A.B}(E,\beta) \approx \frac{16\,\sigma^2\, E\,(E - V_0)}{\left[(\sigma^2 - 1)E + V_0\right]^2}\, e^{-2\kappa_2 a} .\tag{22}$$

If we study, in function of the energy $E$, the factor in front of the damping term $e^{-2\kappa_2 a}$, we note that this factor is maximum for $E = V_0/(1 + \sigma)$. We see that for $\beta = -1/2$, we find the result of the standard tunnel effect through a thick barrier, namely, that the maximum of the factor is obtained for $E = V_0/2$. It is quite plausible to imagine an experiment of tunnel effect where we measure the $E$ value which maximizes this factor, and consequently, we can deduce the correspondent $\beta$ value.

Let us now consider a flat rectangular potential and mass well. If we assume that the discontinuities of the mass and the potential are located at the same positions, it is sufficient to substitute $V_0$ by $-V_0$ in (16). The eigenenergies are the solution of the equation:

$$\cos(p_2\, a/2) = \frac{p_2^2}{p_2^2 \left[1 - (m_1/m_2)^{2\beta+1}\right] + m_2\, V_0\,(m_1/m_2)^{2\beta+1}} ,\tag{23}$$

where $p_2^2 = 2\,m_2\,(E + V_0)/\hbar^2$. We note that if $\beta \neq -1/2$, the eigenvalues depend on the $m_2$ mass of the carrier in the well, which is plausible, but they also are function of the $m_1$ mass of the carrier outside the well, which is, in our opinion, an aberration. On the other hand, if $\beta = -1/2$, eigenvalues are only function of $m_2$ mass of the carrier in the well, besides, these eigenenergies are exactly those of standard spectra of a well with constant $m_2$ mass.

## II. POTENTIAL AND MASS MULTIBARRIER

We consider an alternated series, of period $d$, of potential and masse barriers with thick $a$, and wells with thick $b = d - a$ :

$$V(x) = \begin{cases} V_0 & nd - a/2 \langle x \langle nd + a/2 \\ 0 & nd + a/2 \langle x \langle (n+1)d + a/2 \end{cases} \quad n = 0, 1, 2, ... \quad (24)$$

$$m(x) = \begin{cases} m_2 & nd - a/2 \langle x \langle nd + a/2 \\ m_1 & nd + a/2 \langle x \langle (n+1)d + a/2 \end{cases} \quad n = 0, 1, 2, ... \quad (25)$$

Adjusting the connection rules (3) across the periodical heterojunctions of this medium, we obtain for the quasimomentum $p$ :

$$\cos(p\,d) = \cos(k_1\,b)\cos(k_2\,a) - h(E,\beta)\sin(k_1\,b)\sin(k_2\,a), \quad (26)$$

where $h(E,\beta) = [1 + g(E,\beta)]^{1/2}$. Using the (21) limit, we see that $\lim_{E \to \infty} h(E,\beta) = (\sigma^2 + 1)^2 / 4\sigma^2$. Then, it is clear that $h(E,\beta)$ tends to one only in the $\beta = -1/2$ case, which according to (26), secures that there are no forbidden bands as the energy becomes very large. On the other hand, if $\beta \neq -1/2$, the limit of $h(E,\beta)$ is different from the unity, which involves the existence of forbidden bands even if the energy is very large: this, we think, is a nonphysical result.

## V. CONCLUSION

In this paper, we have studied one dimensional systems with piecewise flat potentials and masses in the shape of a step, a barrier, a well, and a multibarrier. We have applied the connection rules resulting from the Hamiltonian $H = (m^\alpha\,p\,m^\beta\,p\,m^\alpha)/2 + V$, viz., the continuity of $m^\alpha \Psi$ for the wave function, and $m^{\alpha+\beta}\,d\Psi/dx$ for its derivative. We have imposed that the transmission coefficient must tend to the unity as the carrier energy increases indefinitely, which is, in our opinion, very plausible. We have found that the exclusive viable value of $\beta$ parameter is $-1/2$. We thus conclude that the suitable effective Hamiltonian governing one dimensional systems with piecewise constant potentials and masses, must take the form: $H_{flat} = (m^{-1/4}\,p\,m^{-1/2}\,p\,m^{-1/4})/2 + V$. This form implies that the connection rules on the wave function, across abrupt heterojunctions of flat potential and mass, are the continuity of $m^{-1/4}(x)\Psi(x)$ and $m^{-3/4}(x)d\Psi(x)/dx$.